\documentclass[twocolumn,a4paper,10pt]{article}
\usepackage{amsmath,amsfonts}
\usepackage{epsfig}
\usepackage{times}
\usepackage{epsfig}
\usepackage[small,hang]{caption2}
\usepackage{color}
\usepackage{url}
\usepackage{amssymb}

\definecolor{mygray}{gray}{0.6}

\font\smallfont=cmsy10 at 10truept
\mathchardef\bigCircle="280D

\font\bigfont=cmsy10 at 14.4truept
\mathchardef\tiMes="2902        %

\font\Bigfont=cmsy10 at 17.28truept
\mathchardef\DiaMond="2A05        %
\mathchardef\cirCle="2A0E
\mathchardef\BigCircle="2A0D

\font\Bbigfont=cmsy10 at 24.88truept
\mathchardef\buLLet="2B0F

\def\bigCirc{\raise 0.3ex\hbox{$\bigCircle$}\nobreak$\,$}

\def\Bullet{\raise-0.35ex\hbox{$\buLLet$}\nobreak$\,$}

\def\triangledown{\raise 0.2em\hbox{$\bigtriangledown$}\nobreak$\,$}

\def\minisquare{\hbox{${\vcenter{
               \hrule height 0.3pt \kern-0.4pt
               \hbox{\vrule width  0.3pt height 3.0pt \kern 2.6pt
               \vrule width  0.3pt height 3.0pt} \kern-0.4pt
               \hrule height 0.3pt}}$}}
\def\ssquare{\raise 0.175ex\hbox{${\vcenter{
               \hrule height 0.5truept       \kern-0.25truept
               \hbox{\vrule width 0.5truept height 3.0truept \kern 2.75truept
                     \vrule width 0.5truept height 3.0truept} \kern-0.25truept
               \hrule height 0.5truept}}$}\nobreak$\,$}
\def\squarex{\raise 0.175ex\hbox{${\vcenter{
               \hrule height 0.8truept       \kern-1.80truept
          \hbox{\vrule width 0.8truept height 8.0truept \kern-1.95truept
                \raise 0.8truept\hbox{$\tiMes$}     \kern-6.70truept
                \vrule width 0.8truept height 8.0truept} \kern-0.80truept
               \hrule height 0.8truept}}$}\nobreak$\,$}

\def\sqbull{\raise0.175ex\hbox{\vrule height 1.4ex width 1.6ex depth 0.2ex}\nobreak$\,$}
\def\smsqbull{\raise0.175ex\hbox{\vrule height 0.8ex width 0.9ex depth 0.2ex}\nobreak$\,$}

\def\Diamondplus{${\vcenter{\vcenter{\DiaMond} \kern-10truept
                            \hbox{\vrule width .4truept}\kern -3truept
                            \hrule height .4truept}}$\nobreak$\,$}
\newcount\ndots

\def\drawline#1#2{\raise 2.5truept\vbox{\hrule width #1truept height #2truept}}
\def\moonspace#1{\hskip #1truept}

\def\Dashy{\drawline{4.00}{1.00}}     
\def\dashy{\drawline{4.00}{0.75}}     
\def\thindashy{\drawline{4.00}{0.25}}     
\def\dashyspace{\dashy\moonspace{2}}
\def\Dashyspace{\Dashy\moonspace{2}}
\def\thindashyspace{\thindashy\moonspace{2}}
\def\longdashy{\drawline{8.00}{0.75}} 
\def\thinlongdashy{\drawline{8.00}{0.25}} 
\def\longdashyspace{\longdashy\moonspace{2}}
\def\thinlongdashyspace{\thinlongdashy\moonspace{2}}

\def\solid{\drawline{24}{0.75}\nobreak$\,$}

\def\dashbox{\hbox{\dashyspace}}  
\def\Dashbox{\hbox{\Dashyspace}}  
\def\dashed{\hbox {\ndots=0 \loop\ifnum\ndots<3\advance\ndots by 1
        \dashbox\repeat\dashy}\nobreak$\,$}       
\def\Dashed{\hbox {\ndots=0 \loop\ifnum\ndots<3\advance\ndots by 1
        \Dashbox\repeat\Dashy}\nobreak$\,$}       
\def\thindashbox{\hbox{\thindashyspace}}  
\def\thindashed{\hbox {\ndots=0 \loop\ifnum\ndots<3\advance\ndots by 1
        \thindashbox\repeat\thindashy}\nobreak$\,$}       
\def\thindash{\hbox {\ndots=0 \loop\ifnum\ndots<3\advance\ndots by 1
        \thindashbox\repeat\thindashy}\nobreak$\,$}       

\def\longdashbox{\hbox{\longdashyspace}}  
\def\thinlongdashbox{\hbox{\thinlongdashyspace}}  
\def\longdash{\hbox {\ndots=0 \loop\ifnum\ndots<3\advance\ndots by 1
        \longdashbox\repeat\longdashy}\nobreak$\,$}       
\def\thinlongdash{\hbox {\ndots=0 \loop\ifnum\ndots<3\advance\ndots by 1
        \thinlongdashbox\repeat\thinlongdashy}\nobreak$\,$}       

\makeatletter

\newcounter{numbersec}
\renewcommand{\section}[1]{\par\noindent\stepcounter{numbersec}
\par
\vspace{6pt}
\noindent\textbf{\large   \arabic{numbersec} \hspace*{0.3cm} #1 }
\par
\vspace{2pt}
}
\renewcommand{\subsection}[1]{
\par
\vspace{6pt}
\noindent\textbf{#1}
\par
}
\renewcommand{\subsubsection}[1]{%
\par
\vspace{6pt}
\textbf{#1.}
}

\newcommand{\Abstract}{\par\vspace{6pt}\noindent\textbf{\large Abstract}\par\vspace{2pt}}
\newcommand{\Acknowledgments}{\par\vspace{6pt}\noindent\textbf{\large Acknowledgments }\par\vspace{2pt}}

\newenvironment{References}{
\par\vspace{6pt}\noindent\textbf{\large References}\par\vspace{2pt}
\begin{small}\begin{list}{ }{
\itemsep0mm \parsep0mm\labelsep0mm\leftmargin0mm
}}
{\end{list}\end{small}}

\makeatother

\title{\vspace*{-12mm}
\LARGE \sc \textbf{  
Reynolds-number effects in turbulent boundary layers around wing sections
}}
\author{ \Large \bf \textit{ 
R. Vinuesa$^{*}$, P. S. Negi, M. Atzori, A. Hanifi, D. S. Henningson and P. Schlatter}  \\ \\
\bf  \textit{ Linn\'e FLOW Centre, KTH Mechanics, SE-100 44 Stockholm, Sweden} \\
\bf  \textit{ and Swedish e-Science Research Centre (SeRC), Stockholm, Sweden} \\ \\
\underline{\bf $^{*}$rvinuesa@mech.kth.se}
}
\date{}

\begin{document}
%


%

\maketitle
\thispagestyle{empty}



%
%
\Abstract

Four well-resolved LESs of the turbulent boundary layers around a NACA4412 wing section, with $Re_{c}$ ranging from $100,000$ to $1,000,000$, were performed at $5^{\circ}$ angle of attack. By comparing the turbulence statistics with those in ZPG TBLs at approximately matching $Re_{\tau}$, we find that the effect of the adverse pressure gradient (APG) is more intense at lower Reynolds numbers. This indicates that at low $Re$ the outer region of the TBL becomes more energized through the wall-normal convection associated to the APG. This is also reflected in the fact that the inner-scaled wall-normal velocity is larger on the suction side at lower Reynolds numbers. In particular, the wing cases at $Re_{c}=200,000$ and $400,000$ exhibit wall-normal velocities $40\%$ and $20\%$ larger, respectively, than the case with $Re_{c}=1,000,000$. Consequently, the outer-region energizing mechanism associated to the APG is complementary to that present in high-$Re$ TBLs.

%
%
\section{Introduction}

Turbulent boundary layers (TBLs) subjected to adverse pressure gradients (APGs) are relevant for a wide range of industrial applications from diffusers to turbines and wings, and pose a number of open questions regarding their structure and underlying dynamics. The challenges of performing well-resolved simulations of APG TBLs, especially regarding the boundary conditions, were addressed by Spalart and Watmuff (1993). Moreover, as summarized by Monty {\it et al.} (2011), a number of parameters can be used to quantify the magnitude of APGs, a fact that raises serious difficulties when comparing databases from various experimental and numerical sources. In particular, Bobke {\it et al.} (2017) used the Clauser pressure-gradient parameter $\beta=\delta^{*}/\tau_{w} {\rm d}P_{e}/{\rm d}x$ (where $\delta^{*}$ is the displacement thickness, $\tau_{w}$ the wall-shear stress and ${\rm d}P_{e}/{\rm d}x$ the streamwise pressure gradient) to investigate the effect of flow history, {\it i.e.} the impact of the $\beta(x)$ curve on the local features of APG TBLs. They also highlighted the importance of defining cases subjected to a constant value of $\beta$, as also done numerically by Kitsios {\it et al.} (2016), experimentally by Sk\r{a}re and Krogstad (1994), and also in the context of the theory by Mellor and Gibson (1966). Another related work is the study by Vinuesa {\it et al.}~(2017), in which the skin-friction curves of various APG TBLs subjected to different $\beta(x)$ trends were predicted in terms of zero-pressure-gradient (ZPG) data and the averaged pressure-gradient magnitude $\overline{\beta}$. The aim of the present work is to assess the effect of the Reynolds number $Re$ on four APG TBLs subjected to approximately the same $\beta(x)$ distribution. In particular, we consider the turbulent flow around a NACA4412 wing section at four Reynolds numbers based on inflow velocity $U_{\infty}$ and chord length $c$, ranging from $Re_{c}=100,000$ to $1,000,000$. Additional information regarding this database can be found in Vinuesa {\it et al.}~(2018). As discussed by Pinkerton (1938), the NACA4412 wing section is characterized by exhibiting a pressure-gradient distribution essentially independent of $Re$ at moderate angles of attack, a fact that makes this particular airfoil a suitable candidate to study Reynolds-number effects on TBLs given a particular pressure-gradient history. 

\section{Computational setup}

\begin{figure}[h!]
\begin{center}
\includegraphics*[width=0.8\linewidth]{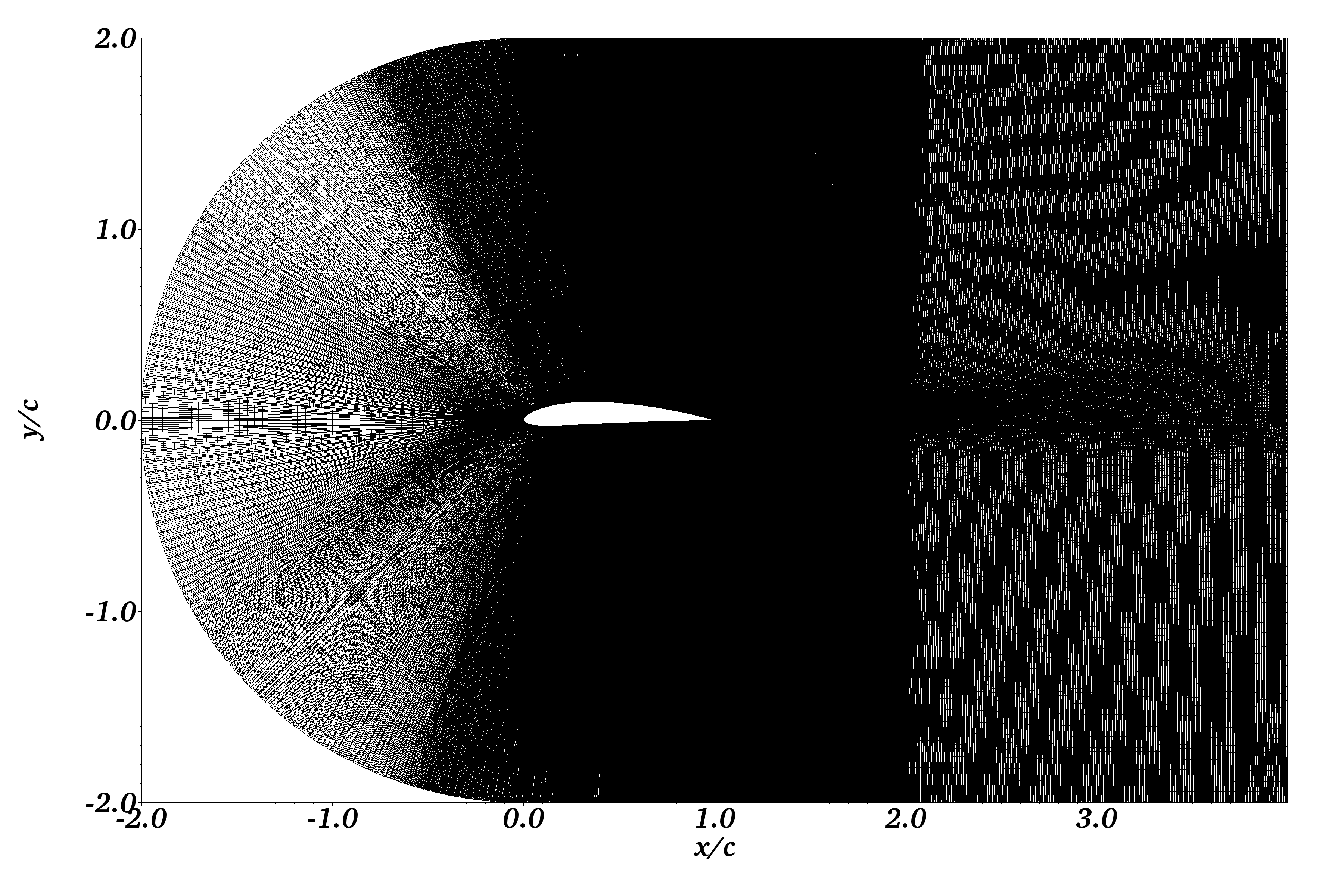}
\includegraphics*[width=0.85\linewidth]{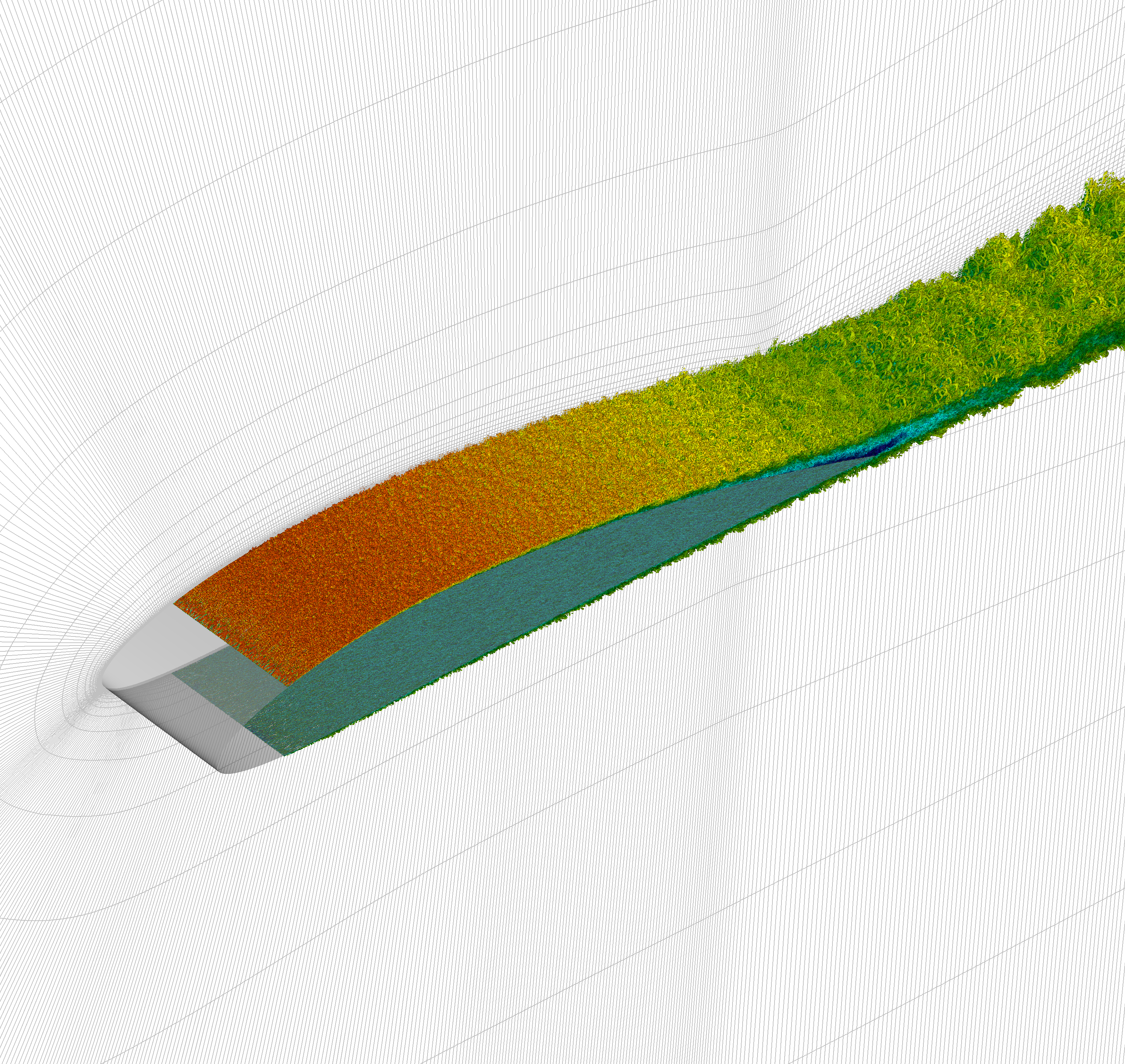}
\caption{\label{setup_fig} (Top) Two-dimensional slice of the spectral-element mesh used in the $Re_{c}=1,000,000$ case, including the indvidual Gauss--Lobatto--Legendre (GLL) points inside the elements. (Bottom) Instantaneous visualization of the coherent vortical structures identified with the $\lambda_{2}$ method in the highest-$Re$ case. The structures are colored by their streamwise velocity, where dark blue denotes $-0.1$ and dark blue 2. A portion of the spectral-element mesh (without GLL points) is also shown.}   
\end{center}
\end{figure}

Well-resolved large-eddy simulations (LESs) of the turbulent flow around a NACA4412 wing section at various Reynolds numbers were carried out using the spectral-element
code Nek5000 (Fischer {\it et al.}, 2008). Additional details regarding the implementation of the spectral-element method (Patera, 1984) in Nek5000 are provided by Deville {\it et al.}~(2002). A total of four Reynolds numbers, namely $Re_{c}=100,000$, $200,000$, $400,000$ and $1,000,000$ (all of them at $5^{\circ}$ angle of attack) is considered in the present study. The highest-$Re$ case is simulated on a computational domain defined by a C-mesh, with streamwise, vertical and spanwise lengths of $L_{x}/c=6$, $L_{y}/c=2$ and $L_{z}/c=0.2$. A precursor RANS simulation of the same geometry was carried out with the $k-\omega$ SST model, and this solution was used as a boundary condition in all boundaries except the outflow (in which a stabilized stress-free condition was employed). Although we observed that for $L_{y}/c \geq 2$ the solution becomes independent of $L_{y}$, we are currently working on non-conformal meshing strategies that will allow us to consider larger domains without the Dirichlet boundary condition from the RANS simulation (Offermans, 2017). Periodicity was imposed in the spanwise direction. The resolution in the $Re_{c}=1,000,000$ case follows these guidelines around the wing section: $\Delta x^{+} < 27$, $\Delta y ^{+}_{w}<0.96$ (spacing of the first grid point in the wall-normal direction) and $\Delta z^{+}<13.5$ (where the superscript `+' denotes viscous scaling in terms of the friction velocity $u_{\tau}$ and the viscous length $\ell^{*}=\nu / u_{\tau}$). An additional condition based on the Kolmogorov scale $\eta= \left (\nu^{3} / \varepsilon \right )^{1/4}$, where $\varepsilon$ is the local isotropic dissipation, was defined for the wake: $\Delta x /\eta < 13.5$. A total of 4.5 million spectral elements was used to discretize the domain with a polynomial order $N=7$, which amounts to a total of 2.28 billion grid points. In Figure~\ref{setup_fig}~(top) we show a two-dimensional slice of the spectral-element mesh used to discretize this case. The numerical setup is similar to the one employed in the direct numerical simulation (DNS) by Vinuesa {\it et al.} (2017) at $Re_{c}=400,000$, and the relaxation-term (RT) filter approach by Schlatter {\it et al.} (2004) was used to perform the LES. Note that the RT filter implemented in Nek5000 was thoroughly validated by Negi {\it et al.}~(2018) using the DNS at $Re_{c}=400,000$ as a benchmark. An excellent agreement in mean flow, Reynolds-stress tensor and turbulent kinetic energy (TKE) budgets around the wing was reported. An instantaneous visualization of the flow around the wing section in the $Re_{c}=1,000,000$ case is shown in Figure~\ref{setup_fig}~(bottom), where the level of detail achieved in the simulation can be observed.

\section{Results}

\begin{figure}[h!]
\begin{center}
\includegraphics*[width=0.82\linewidth]{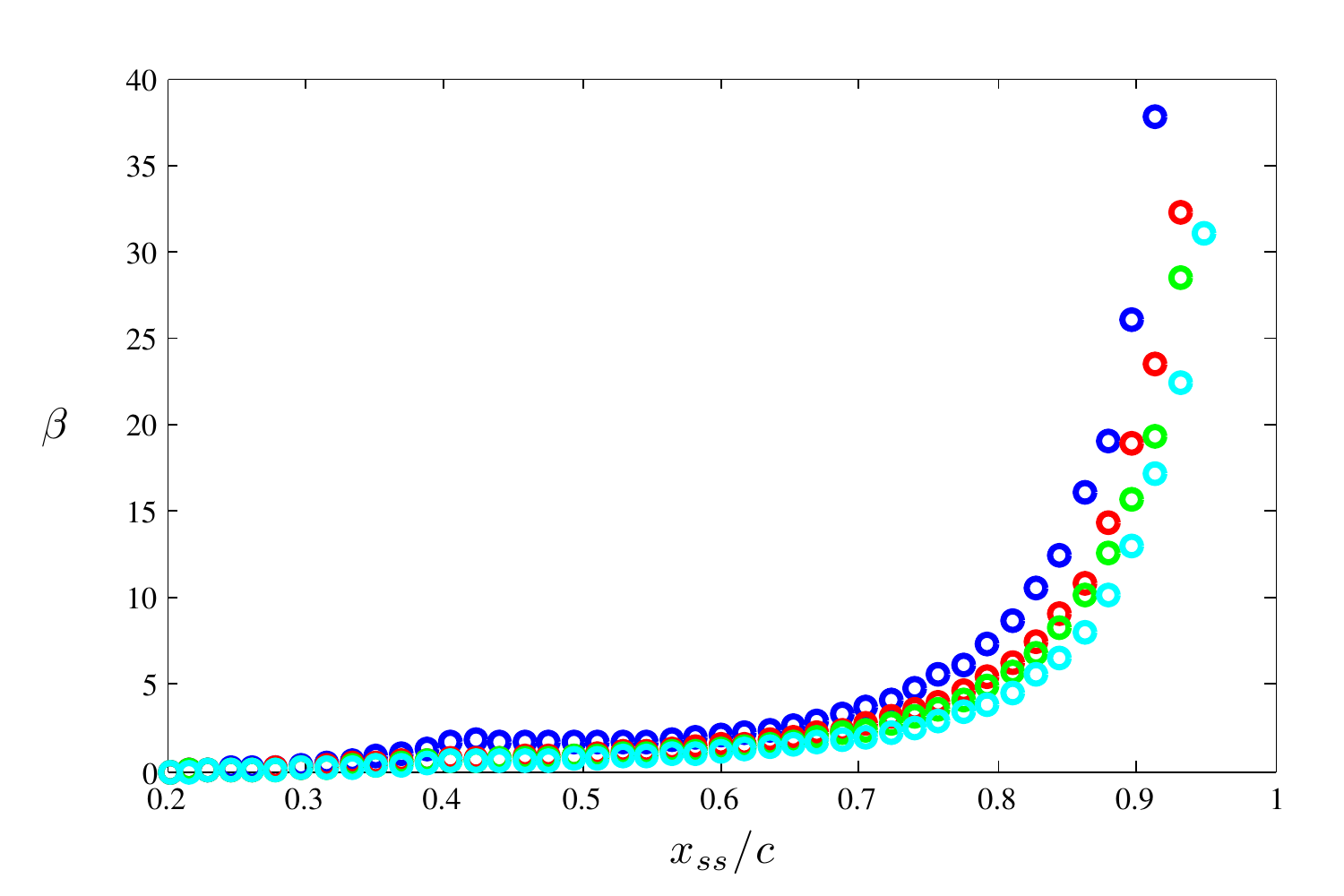}
\caption{\label{beta_fig} Streamwise evolution of $\beta$ on the suction side of the four wings, where: $(\textcolor{blue}{\circ})$ $Re_{c}=100,000$, $(\textcolor{red}{\circ})$ $Re_{c}=200,000$, $(\textcolor{green}{\circ})$ $Re_{c}=400,000$ and $(\textcolor{cyan}{\circ})$ $Re_{c}=1,000,000$.}   
\end{center}
\end{figure}

In Figure~\ref{beta_fig}, the streamwise evolution of the Clauser pressure-gradient parameter $\beta$ is shown on the suctions side (denoted by $ss$)  for the four wing cases under consideration. It can be observed that for $Re_{c} \geq 200,000$ the $\beta(x)$ curves are approximately the same with only small relative differences (of around $10\%$) beyond $x_{ss}/c \simeq 0.9$. This observation is in agreement with Pinkerton~(1938). The different behavior observed at $Re_{c}=100,000$ is associated to the very low Reynolds number combined with the strong APG conditions. Thus, these simulations allow us to study the impact of $Re$ on three APG TBLs with approximately the same pressure-gradient history. As discussed by Bobke {\it et al.}~(2017), the state of a TBL is not uniquely determined by the APG magnitude, but rather by the accumulated pressure-gradient effect, {\it i.e.}, by the $\beta(x)$ curve. Streamwise APGs lead to an increased wall-normal convection, a fact that increases the boundary-layer thickness and produces a more prominent outer region. This leads to more energetic large-scale motions, and to reduced wall-shear stress. One of the goals of this study is to assess the effect of APG and of $Re$ on the outer region of the TBLs, and to try to identify their complementing energizing effects on the large-scale motions.

\begin{figure}[h!]
\begin{center}
\includegraphics*[width=0.85\linewidth]{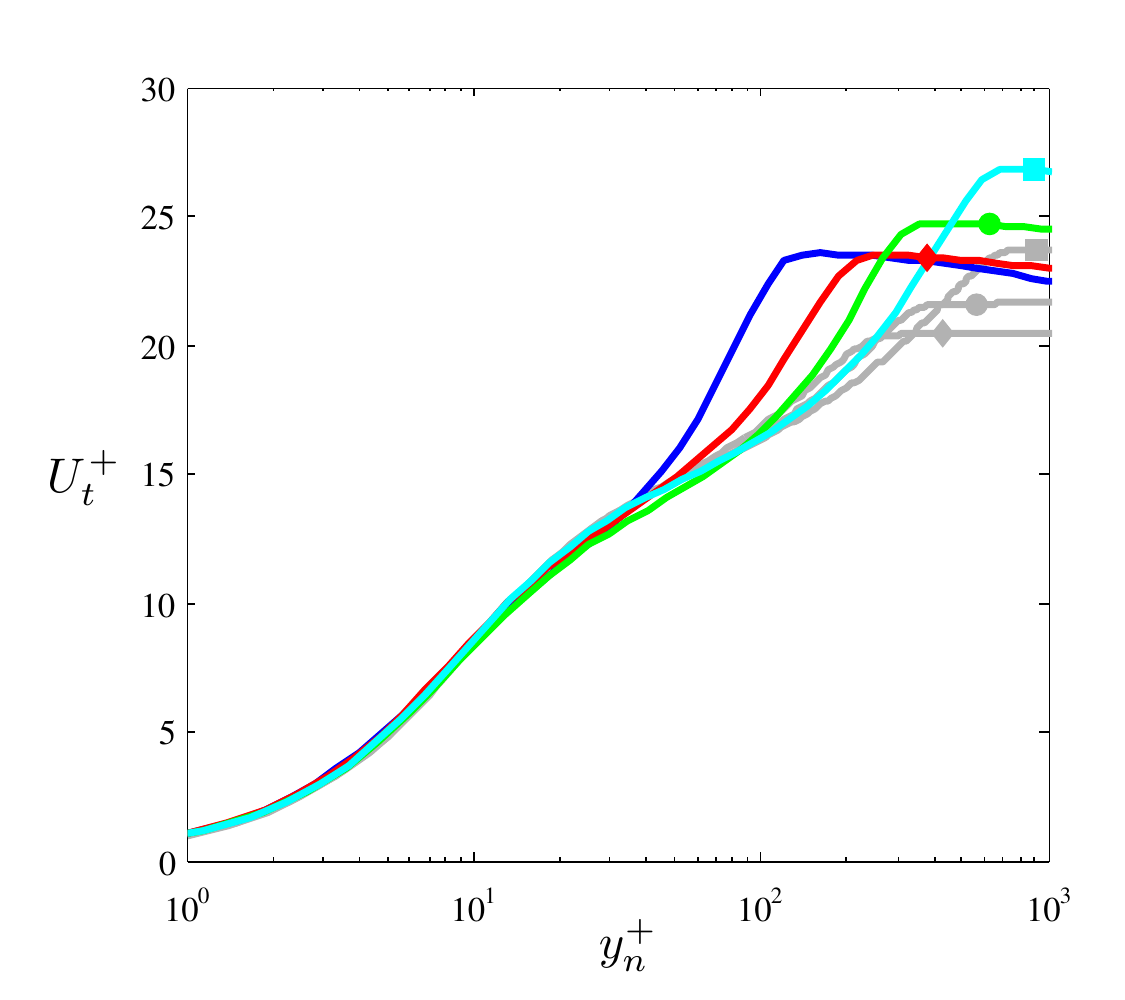}
\includegraphics*[width=0.85\linewidth]{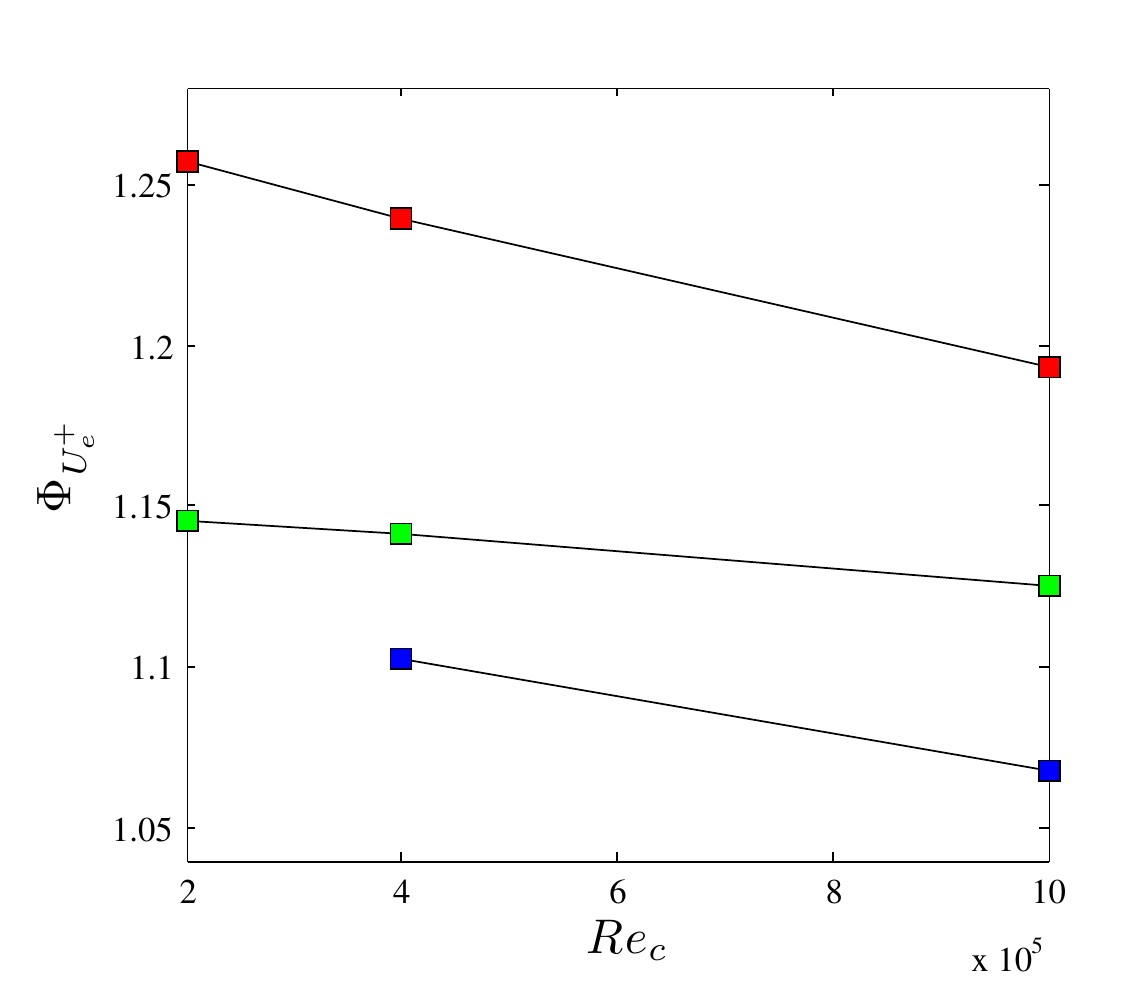}
\includegraphics*[width=0.85\linewidth]{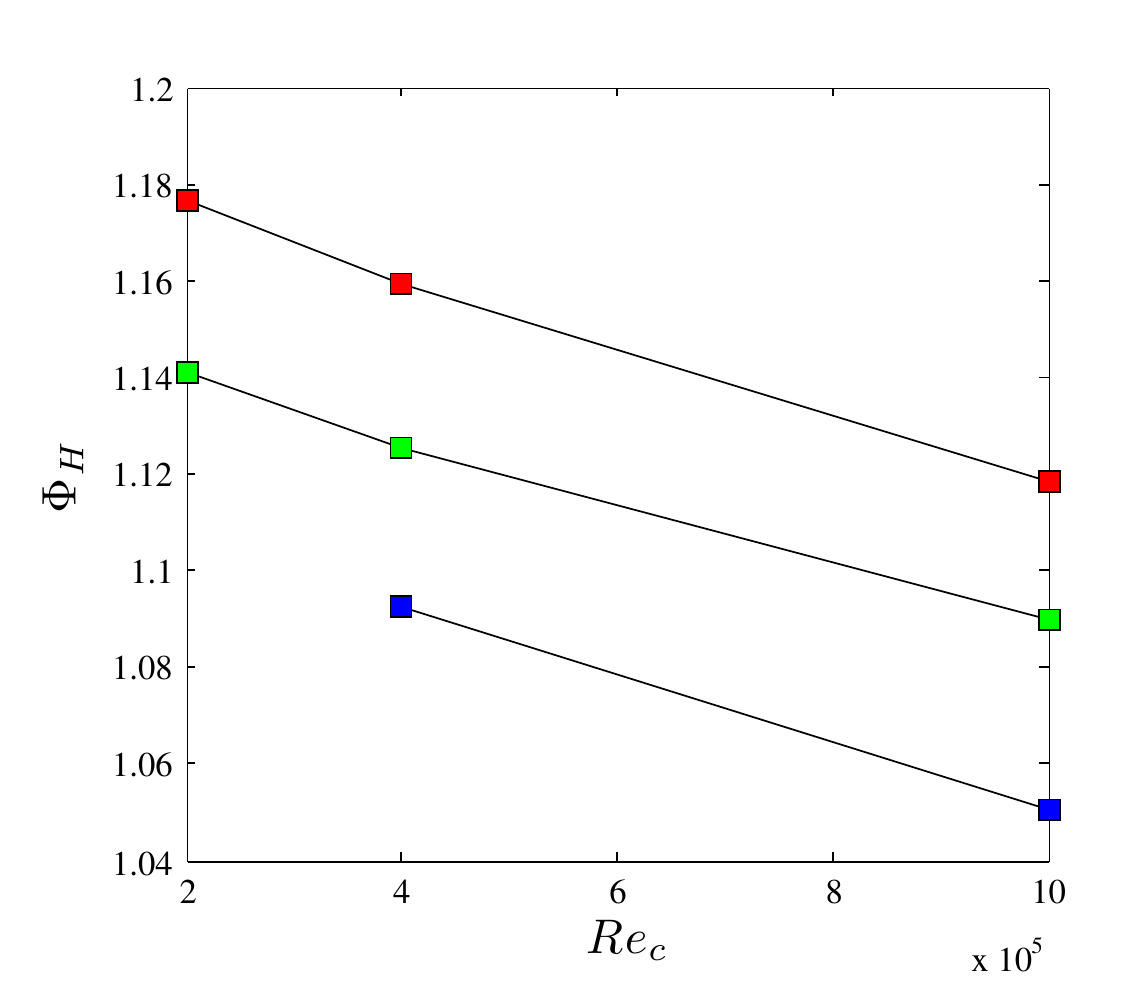}
\caption{\label{Up_fig} (Top) Inner-scaled tangential mean velocity profiles at $x_{ss}/c=0.6$ for the four wing cases under study. Colors as in Figure~\ref{beta_fig}, and {\color{mygray}\solid} denotes ZPG TBL data at matching $Re_{\tau}$. The matched $U^{+}_{t}$ profiles for W10 and ZPG10 are denoted by $\left ( \blacksquare \right )$, for W4 and ZPG4 by $\left ( \bullet \right )$ and for W2 and ZPG2 by $\left ( \blacklozenge \right )$. (Middle) Ratio of $U^{+}_{e}$ and (bottom) $H$ between wing and ZPG at matching $Re_{\tau}$, where (\textcolor{blue}{$\blacksquare$}), (\textcolor{green}{$\blacksquare$}) and (\textcolor{red}{$\blacksquare$}) denote ratios at $x_{ss}/c=0.4$, $0.6$ and $0.7$, respectively.}   
\end{center}
\end{figure}

The inner-scaled mean velocity profile at $x_{ss}/c=0.6$ is compared among the four wing cases, and with ZPG TBL results from DNS (Schlatter and \"Orl\"u, 2010) at approximately matching $Re_{\tau}$ (which is the friction Reynolds number, defined in terms of the friction velocity $u_{\tau}$ and the $99\%$ boundary-layer thickness $\delta_{99}$), in Figure~\ref{Up_fig}~(top). The value of $\delta_{99}$ is calculated following the method by Vinuesa {\it et al.}~(2016). This figure shows the effect of the moderate local APG, given by $\beta \simeq 1.3$ (for $Re_{c} \geq 200,000$), which produces larger values of the inner-scaled edge velocity $U^{+}_{e}$ ({\it i.e.}, lower skin friction) and reduced inner-scaled mean velocities in the buffer region with respect to the corresponding ZPG cases. The integral parameters of the various wing cases and the corresponding ZPG TBLs are reported in Table~\ref{table_wings}. The values of $Re_{\tau}$ are 209, 329 and and 600 for the wings at $Re_{c}=200,000$, $400,000$ and $1,000,000$, respectively. The friction Reynolds numbers in the corresponding ZPG cases are very close, although for instance the value of $Re_{\tau}$ closest to that in the $Re_{c}=200,000$ case was somewhat higher: 252. The APG increases the boundary-layer thickness through wall-normal convection, a fact that is manifested in the larger values of $Re_{\theta}$ (Reynolds number based on momentum thickness) with respect to the ZPG profiles at matched $Re_{\tau}$. This is also reflected in the larger values of the shape factor $H=\delta^{*}/\theta$ (which is the ratio of the displacement and the momentum thicknesses), which are due to the boundary-layer thickening associated to the APG. This thickening is connected to a lower velocity gradient at the wall, {\it i.e.} a reduced wall-shear stress, which is consistent with the lower skin-friction coefficients $C_{f}=2 \left (u_{\tau} / U_{e} \right )^{2}$ observed in the wing profiles at $x_{ss}/c=0.6$. It is possible to further quantify the effect of the APG by comparing the difference between some of these integral parameters in the wing and in the ZPG at approximately the same $Re_{\tau}$. In this study, we consider the parameter $\Phi_{U^{+}_{e}}$, which is defined as the ratio between the $U^{+}_{e}$ from the TBL on the wing at a certain $Re_{c}$ and $x_{ss}$, divided by the same quantity in a ZPG TBL with the same $Re_{\tau}$. Since one of the effects of the APG is to increase $U^{+}_{e}$, the value of $\Phi_{U^{+}_{e}}$ should in principle be larger than 1 in APGs. In Figure~\ref{Up_fig}~(middle) we show the evolution of $\Phi_{U^{+}_{e}}$ at $x_{ss}/c=0.6$ as a function of $Re_{c}$, in the three wing cases for which we have matched-$Re_{\tau}$ ZPGs, namely from $Re_{c}=200,000$ to $1,000,000$. Not only this quantity is larger than 1 over the whole $Re$ range, but it shows a decreasing trend with increasing Reynolds number, from a maximum value of around 1.15 at $Re_{c}=200,000$ to a value of 1.13 at $Re_{c}=1,000,000$. Note that the value at $Re_{c}=400,000$ is only slightly below the one at $Re_{c}=200,000$, which is connected to the fact that the $Re_{\tau}$ from ZPG is slightly larger than that in the wing in this case. In any case, the decreasing trend with $Re_{c}$ is also observed at $x_{ss}/c=0.4$ and 0.7, as also shown in Figure~\ref{Up_fig}~(middle). This is noteworthy since it suggests that the effect of the APG, given approximately the same $\beta(x)$ curve, is more intense at lower Reynolds numbers. In Figure~\ref{Up_fig}~(bottom) we show another indicator, namely the ratio between the shape factor in the wing and in a ZPG at matched $Re_{\tau}$, which we denote by $\Phi_{H}$. This quantity is also larger in APGs than in ZPGs, due to the fact that the streamwise deceleration of the TBL increases the boundary-layer thickness, as discussed above. Furthermore, this quantity also decreases with $Re$ at $x_{ss}/c=0.6$, with values ranging from 1.14 to 1.09. This behavior is also consistent with the trend observed for $x_{ss}/c=0.4$ and 0.7. It can be therefore concluded that APG effects on the mean flow appear to be stronger in lower-$Re$ TBLs, an observation also made in the recent experimental study by Sanmiguel Vila {\it et al.}~(2017).
\begin{table*}
\begin{center}
\def~{\hphantom{0}}
\begin{tabular}{c c c c c c c c}
Parameter & W1 &  W2 & W4 &  W10 & ZPG2 & ZPG4 & ZPG10 \\[3pt]
\hline
$\beta$	  & 2.1 & 1.4& 1.3 & 1.1 &	$\simeq 0$ & $\simeq 0$ & $\simeq 0$  \\
$Re_{\tau}$	  & 120 & 209& 329 & 600& $252$ & $359$ & 671 \\
$Re_{\theta}$	  & 445 & 727& $1,208$ & $2,350$ &	$678$ & 	$1,007$ & $2,001$  \\
$C_{f}$	  & $3.7 \times 10^{-3}$ & $3.6 \times 10^{-3}$ & $3.3 \times 10^{-3}$ &  $2.8 \times 10^{-3}$ &	$4.8 \times 10^{-3}$ & $4.3 \times 10^{-3}$ & $3.5 \times 10^{-3}$  \\
$H$	  & 1.90 & 1.68 & 1.63 & 1.54 &	$1.47$ & $1.45$ & 1.41 \\
\end{tabular}
\caption{Boundary-layer parameters at $x_{ss}/c=0.6$ for the four cases (denoted by W1--W10 for increasing $Re_{c}$). ZPG2, ZP4 and ZPG10 denote the DNS ZPG TBL case (Schlatter and \"Orl\"u, 2010) approximately matching the $Re_{\tau}$ of the W2, W4 and W10 wing cases, respectively.}   
\label{table_wings}
\end{center}
\end{table*}

\begin{figure}[h!]
\begin{center}
\includegraphics*[width=0.85\linewidth]{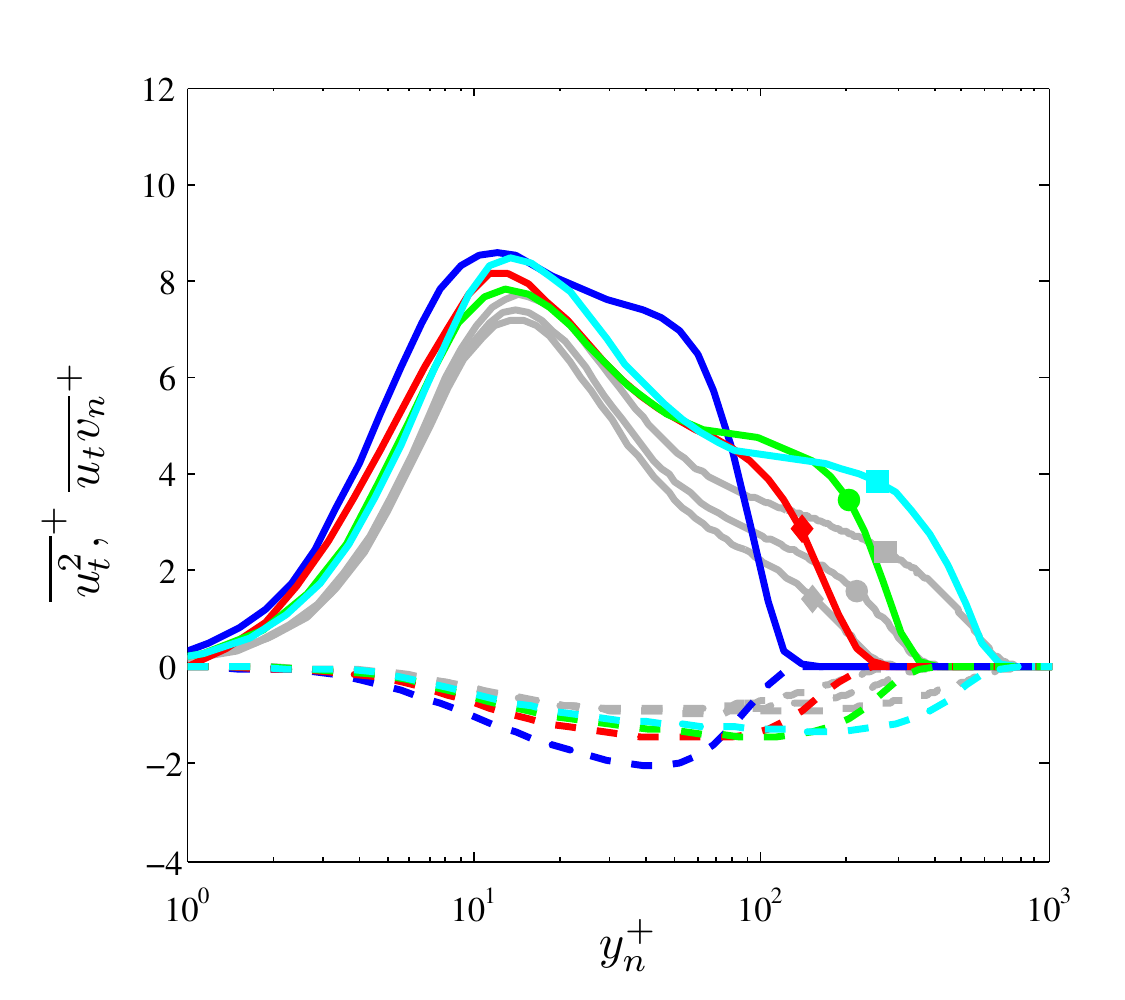}
\includegraphics*[width=0.85\linewidth]{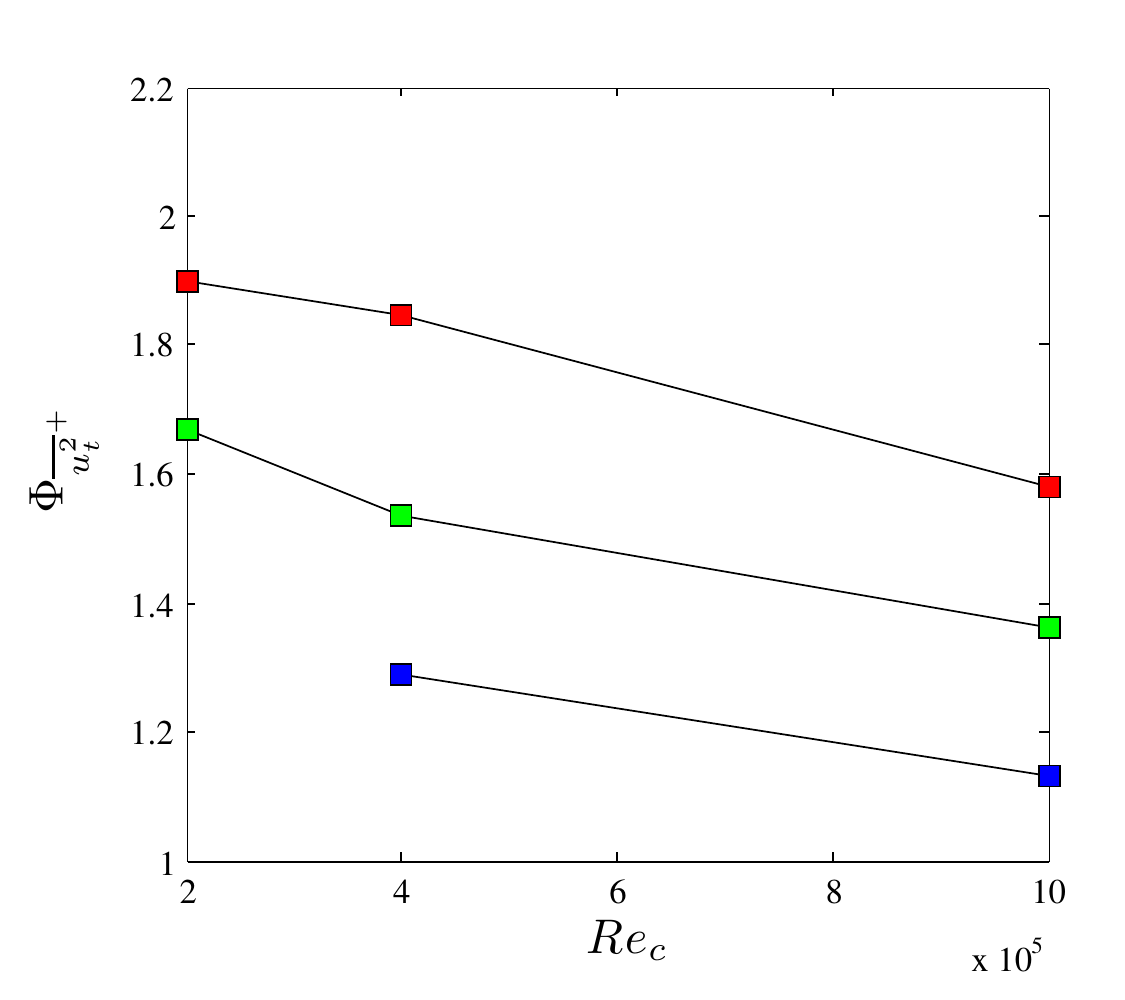}
\includegraphics*[width=0.85\linewidth]{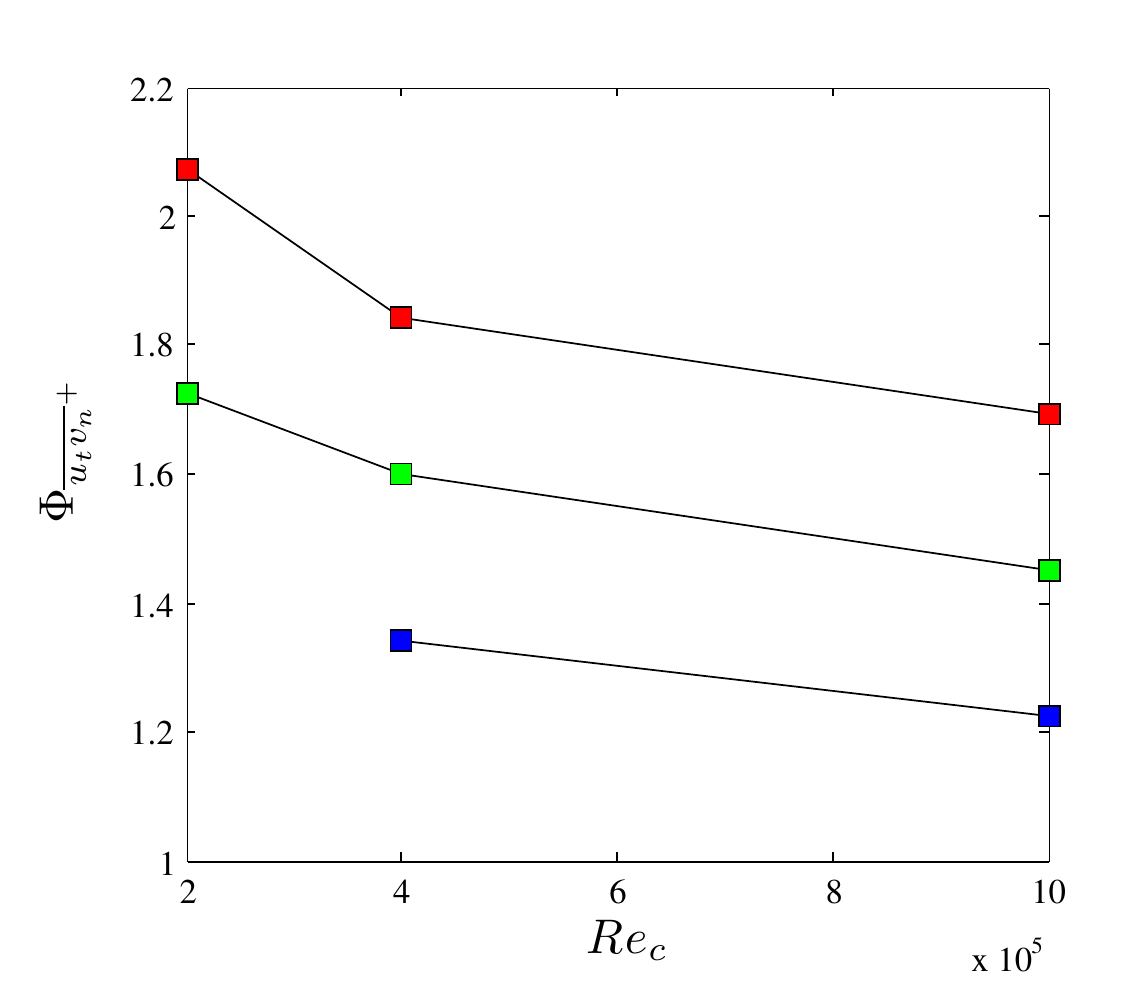}
\caption{\label{uup_fig} (Top) Inner-scaled tangential velocity fluctuations and Reynolds-shar stress at $x_{ss}/c=0.6$ for the four wing cases under study. (Middle) Ratio of $\overline{u^{2}_{t}}^{+}$ and (bottom) $\overline{u_{t}v_{n}}^{+}$ between wing and ZPG at $y_{n}/\delta_{99} \simeq 0.2$ for matching $Re_{\tau}$. Legend as in Figure~\ref{Up_fig}.}   
\end{center}
\end{figure}

Further insight regarding the APG effects (given the same streamwise flow history) on TBLs at various $Re$ can be gained by analyzing some componets of the Reynolds-stress tensor. In particular, we focus on the inner-scaled tangential velocity fluctuations $\overline{u^{2}_{t}}^{+}$ and the Reynolds-shear stress $\overline{u_{t}v_{n}}^{+}$. These quantities are shown in Figure~\ref{uup_fig}~(top) for the four wing cases at $x_{ss}/c=0.6$, together with the corresponding ZPG cases at matched $Re_{\tau}$ as summarized in Table~\ref{table_wings}. In this figure all the quantities are expressed in terms of the local directions tangential ($t$) and normal ($n$) to the wing surface. As discussed for instance by Harun {\it et al.}~(2013), the APG leads to more energetic large-scale motions, which produce larger energy concentration in the TBL outer region. This is associated with the emergence of an outer peak in the stremawise velocity fluctuation profile, for large enough values of $\beta$. Harun {\it et al.}~(2013) also showed that the larger energy concentration in the outer region for APGs is not a consequence of the inner scaling, since the outer-scaled streamwise velocity fluctuation profiles also show an increasing outer peak for higher $\beta$ (and although the near-wall peak increases in inner scaling with $\beta$, it actually decreases when scaled in outer units). This observation suggests that the large-scale motions induced by the APG may be different from those arising in high-$Re$ ZPG TBLs. On the one hand, they appear to be more energetic, and on the other hand they are larger in $y$ and more inclined with respect to the wall, as reported by Maciel {\it et al.}~(2017) in their numerical work on APG TBLs. Note that they identified three-dimensional coherent structures in the instantaneous fields corresponding to intense Reynolds-shear events and reported the probability density function of their sizes. Figure~\ref{uup_fig}~(top) clearly reflects the larger energy accumulation in the outer region of the APG TBLs, both in the tangential velocity fluctuations and in the Reynolds-shear stresses. Interestingly, the difference between the values in the outer region of both profiles and the corresponding ZPG data at matched $Re_{\tau}$ also appears to become smaller as $Re$ is increased. As done before for the mean flow, we define the variable $\Phi_{\overline{u^{2}_{t}}^{+}}$, which is the ratio between the tangential velocity fluctuations in the wing at a wall-normal distance of $y_{n}/\delta_{99} \simeq 0.2$, and those in the ZPG TBL at matched $Re_{\tau}$. This wall-normal location is chosen because it approximately defines the end of the overlap region in the current Reynolds-number range. Figure~\ref{uup_fig}~(middle) shows the evolution with $Re_{c}$ of $\Phi_{\overline{u^{2}_{t}}^{+}}$, which is also decreasing, and ranges from around 1.67 at $Re_{c}=200,000$ to 1.36 at $Re_{c}=1,000,000$. The evolution observed at $x_{ss}/c=0.6$ is consistent with the curves for $x_{ss}/c=0.4$ and 0.7, in agreement with the mean-flow ratios discussed in Figure~\ref{Up_fig}. Moreover, an analogous ratio, $\Phi_{\overline{u_{t}v_{n}}^{+}}$, is defined for the Reynolds-shear stress and is shown for the same profiles in Figure~\ref{uup_fig}~(bottom). This quantity also shows a decreasing trend with $Re$. These are all consistent with the statement that APG effects are stronger at lower $Re$,  whereas at higher Reynolds number the relative effect with respect to a ZPG TBL is smaller. 

\begin{figure}[h!]
\begin{center}
\includegraphics*[width=0.82\linewidth]{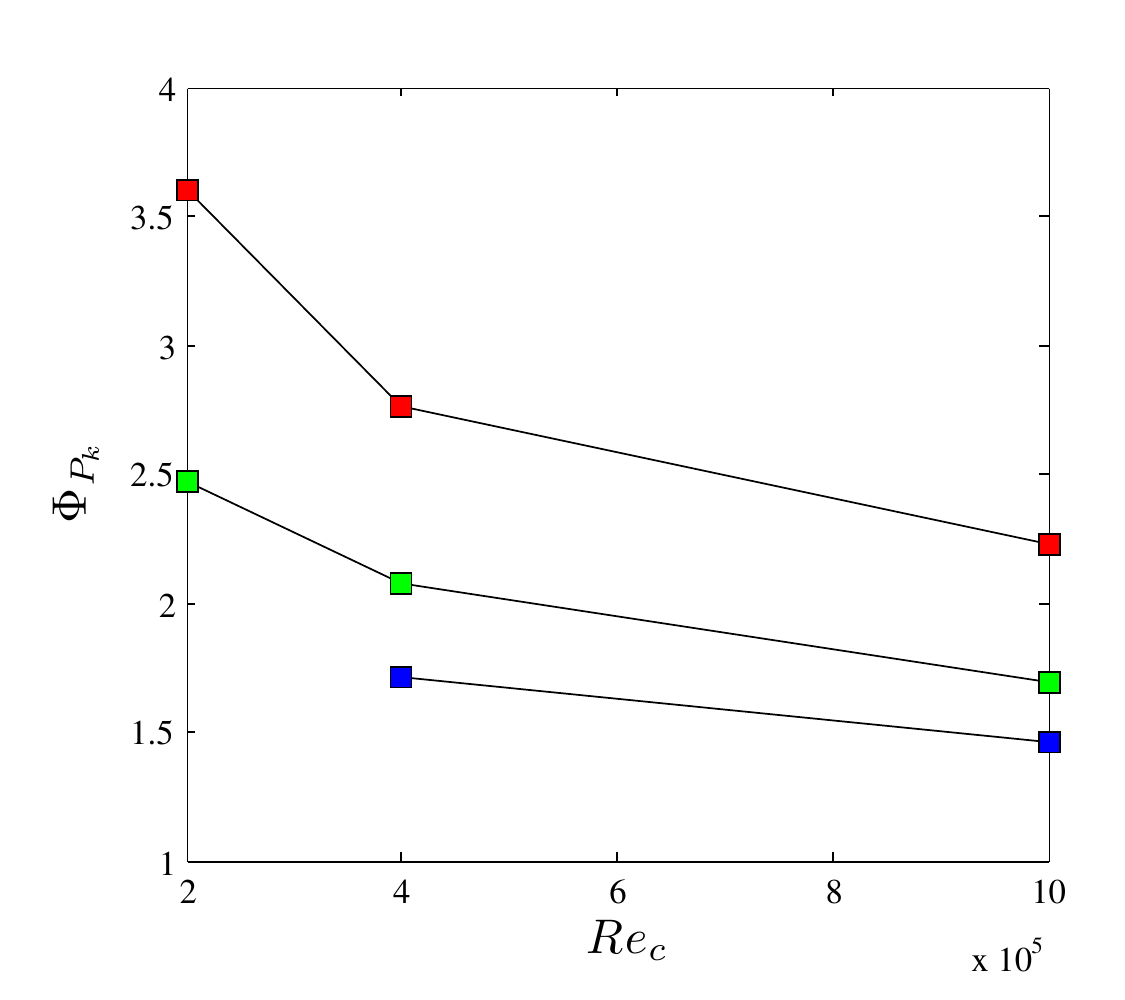}
\caption{\label{Prod_fig} Ratio of the TKE production between wing and ZPG at $y_{n}/\delta_{99} \simeq 0.2$ for matching $Re_{\tau}$. Legend as in Figure~\ref{Up_fig}.}   
\end{center}
\end{figure}

The results above show that the mechanisms producing a more energetic outer region in APG TBLs are complementary to those present in high-$Re$ ZPG TBLs. Moreover, the large-scale motions resulting from increases either in $\beta$ and in $Re$ are different, as also indicated by Maciel {\it et al.}~(2017). It is interesting to note that the discrepancy between APG and ZPG TBLs, for profiles with the same range of scales ({\it i.e.}, at the same $Re_{\tau}$), decreases with increasing Reynolds number. This shows that low-$Re$ TBLs are more sensitive to the effect of APGs, namely the increase in wall-normal convection and the reduction of mean velocity gradient at the wall. An additional indication of this effect can be observed in Figure~\ref{Prod_fig}, which shows the ratio $\Phi_{P_{k}}$ as a function of $Re_{c}$ at the three previous streamwise locations, namely $x_{ss}/c=0.4$, 0.6 and 0.7. This ratio is defined similarly to the ones discussed above, and in this case we show the TKE production $P_{k}$ at $y_{n}/\delta_{99} \simeq 0.2$. This ratio exhibits a trend consistent with the ones observed in Figures~\ref{Up_fig} and \ref{uup_fig}, and shows that the outer-region production is larger at low Reynolds numbers, and decreases with $Re$. This observation is connected to the higher sensitivity to APG effects observed at low Reynolds numbers, which in turn is consistent with the differences in the large-scale motions present in APGs. In particular, the ratio $\Phi_{P_{k}}$ decreases from around 2.47 at $Re_{c}=200,000$ to 1.69 at $Re_{c}=1,000,000$, for $x_{ss}/c=0.6$.  

\begin{figure}[h!]
\begin{center}
\includegraphics*[width=0.82\linewidth]{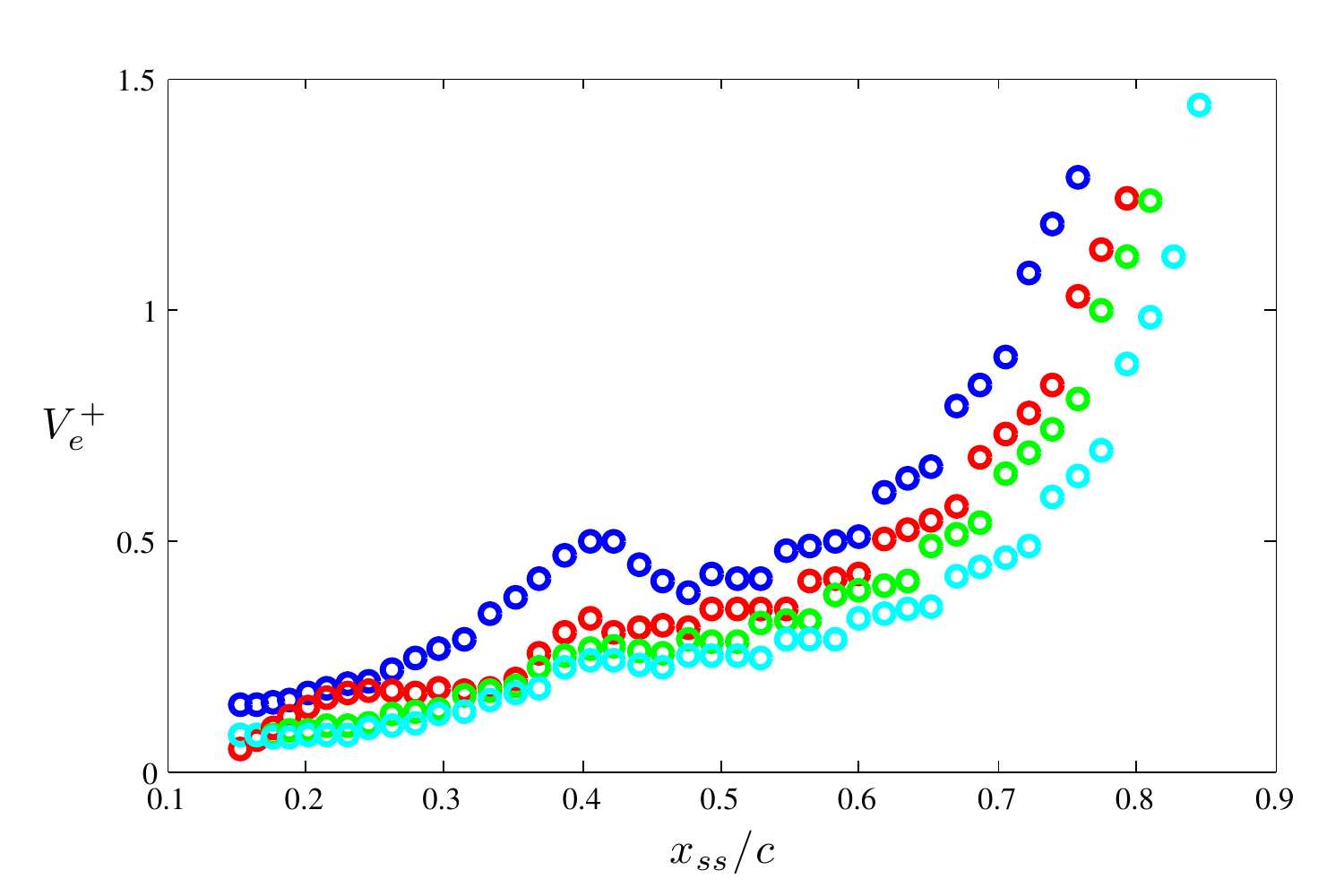}
\caption{\label{Vep_fig} (Left) Streamwise evolution of the inner-scaled wall-normal velocity at the boundary-layer edge for the four wing cases under study. Colors as in Figure~\ref{beta_fig}.}   
\end{center}
\end{figure}

One explanation for the higher sensitivity to APGs at low $Re$ can be obtained from the wall-normal convection: larger wall-normal velocities produce a more prominent development of the outer region, therefore promoting the development of very energetic large-scale motions. Figure~\ref{Vep_fig} shows the streamwise evolution, on the suction side, of the inner-scaled wall-normal velocity at the boundary-layer edge, for the four simulations. The boundary-layer edge is a sensitive quantity, and the problematic associated to its determination in PG TBLs is discussed by Vinuesa {\it et al.}~(2016) and Coleman {\it et al.}~(2018); this sensitivity is reflected in the shape of the curves. This figure shows that the wall-normal velocity is higher, throughout the whole suction side, at lower Reynolds numbers. In particular, the $V^{+}_{e}$ values are, on average, around $40\%$ larger at $Re_{c}=200,000$ than in the highest-$Re$ wing case. Regarding the simulation at $Re_{c}=400,000$, the inner-scaled wall-normal velocity is around $20\%$ larger than at $Re_{c}=1,000,000$. This again supports the statement that APG effects are stronger at low $Re$, which in turn produce a more energetic outer region compared with that at higher $Re$.

\section{Conclusions}

The present contribution summarizes the results of four well-resolved LESs of the flow around a NACA4412 wing section, with $Re_{c}$ ranging from $100,000$ to $1,000,000$, all of them at $5^{\circ}$ angle of attack. We compare the turbulence statistics on the suction side, with focus on the profiles at $x_{ss}/c=0.6$, with ZPG TBL results at approximately matching $Re_{\tau}$. We quantified the difference between a particular wing case and the corresponding ZPG with matched $Re_{\tau}$ through ratios of $U^{+}_{e}$ and $H$ (to characterize mean flow quantities), and of $\overline{u^{2}_{t}}^{+}$ and $\overline{u_{t} v_{n}}^{+}$ at $y_{n} / \delta_{99} \simeq 0.2$ (to analyze some components of the Reynolds stresses in the outer region). All these ratios are larger than 1, which is consistent with the effect of a streamwise adverse pressure gradient. Interestingly, these ratios show a decreasing trend with $Re$, indicating that the APG effects  are more intense at lower Reynolds numbers, in particular when it comes to energizing the outer region. The APG increases the wall-normal convection, which results in a thicker boundary layer and a larger outer region. This allows the development of more energetic large-scale motions, and constitutes a mechanism different, although complementary to, that present at high-$Re$ TBLs. This conclusion is supported by the evolution of the inner-scaled wall-normal velocity, which also decrease with Reynolds numbers, and further indicates that the outer-region energy responds differently to APGs with varying Reynolds number.

\Acknowledgments

The simulations were performed on resources provided by the Swedish National Infrastructure for Computing (SNIC) at the Center for Parallel Computers (PDC), in Stockholm (Sweden), and by the Partnership for Advanced Computing in Europe (PRACE) at the Barcelona Supercomputing Center (BSC) in Barcelona (Spain). 
RV and PS acknowledge the funding provided by the Swedish Research Council (VR) and from the Knut and Alice Wallenberg Foundation. This research is also supported by the ERC Grant No. ``2015-AdG-694452, TRANSEP'' to DH.

%
\begin{References}

\item Spalart, P. R. and Watmuff, J. H. (1993), Experimental and numerical study of a turbulent boundary layer with pressure gradients, {\it J. Fluid Mech.} Vol. 249, pp. 337--371.

\item Monty, J. P., Harun, Z. and Marusic, I. (2011), A parametric study of adverse pressure gradient turbulent boundary layers, {\it Int. J. Heat Fluid Flow} Vol. 32, pp. 575--585.

\item Bobke, A., Vinuesa, R., \"Orl\"u, R. and Schlatter, P. (2017), History effects and near equilibrium in adverse-pressure-gradient turbulent boundary layers, {\it J. Fluid Mech.} Vol. 820, pp. 667--692.

\item Kitsios, V., Atkinson, C., Sillero, J. A., Borrell, G.,  Gungor, A. G., Jim\'enez, J. and Soria, J.  (2016), Direct numerical simulation of a self-similar adverse pressure gradient turbulent boundary layer, {\it Int. J. Heat Fluid Flow} Vol. 61, pp. 129--136.

\item Sk\r{a}re, P. E. and Krogstad, P. \r{A}. (1994), A turbulent equilibrium boundary layer near separation, {\it J. Fluid Mech.} Vol. 272, pp. 319--348.

\item Mellor, G. L. and Gibson, D. M. (1966), Equilibrium turbulent boundary layers, {\it J. Fluid Mech.} Vol. 24, pp. 225--253.

\item Vinuesa, R., \"Orl\"u, R., Sanmiguel Vila, C., Ianiro, A., Discetti, S. and Schlatter, P. (2017), Revisiting history effects in adverse-pressure-gradient turbulent boundary layers, {\it Flow Turbul. Combust.} Vol. 99, pp. 565--587.

\item Vinuesa, R., Negi, P. S., Atzori, M., Hanifi, A., Henningson, D. S. and Schlatter, P. (2018), Turbulent boundary layers around wing sections up to $Re_{c}=1,000,000$, {\it Int. J. Heat Fluid Flow}, Vol. 72, pp. 86--99.

\item Pinkerton, R. M. (1938),  The variation with Reynolds number of pressure distribution over an airfoil section, {\it NACA Annual Report} Vol.  24, pp. 65--84.

\item Fischer, P. F., Lottes, J. W. and Kerkemeier, S. G. (2008), NEK5000: Open source spectral element CFD solver. Available at: \url{http://nek5000.mcs.anl.gov}

\item Patera, A. T. (1984), A spectral element method for fluid dynamics: laminar flow in a channel expansion, {\it J. Comput. Phys.} Vol. 54, pp. 468--488.

\item Deville, M. O., Fischer, P. F. and Mund, E. H. (2002), High-order methods for incompressible fluid flow, {\it Cambridge University press}, Cambridge.

\item Offermans, N. (2017), Towards adaptive mesh refinement in Nek5000 {\it Licentiate Thesis}, KTH Royal Institute of Technology, Stockholm, Sweden.

\item Vinuesa, R., Hosseini, S. M., Hanifi, A., Henningson, D. S. and Schlatter, P. (2017), Pressure-gradient turbulent boundary layers developing around a wing section, {\it Flow Turbul. Combust.} Vol. 99, pp. 613--641.

\item Schlatter, P., Stolz, S. and Kleiser, L. (2004), LES of transitional flows using the approximate deconvolution model, {\it Int. J. Heat Fluid Flow} Vol. 25, pp. 549--558.

\item Negi, P. S., Vinuesa, R., Hanifi, A., Schlatter, P. and  Henningson, D. S. (2018), Unsteady aerodynamic effects in small-amplitude pitch oscillations of an airfoil {\it Int. J. Heat Fluid Flow} Vol. 71, pp. 378--391.

\item Schlatter, P. and \"Orl\"u, R. (2010), Assessment of direct numerical simulation data of turbulent boundary layers, {\it J. Fluid Mech.} Vol. 659, pp. 116--126.

\item Vinuesa, R., Bobke, A., \"Orl\"u, R. and Schlatter, P. (2016), On determining characteristic length scales in pressure-gradient turbulent boundary layers. {\it Phys. Fluids}, Vol. 28, pp. 055101.

\item Sanmiguel Vila, C., \"Orl\"u, R., Vinuesa, R., Schlatter, P., Ianiro, A., and Discetti, S. (2017), Adverse-pressure-gradient effects on turbulent boundary layers: statistics and flow-field organization. {\it Flow Turbul. Combust.}, Vol. 99, pp. 589--612.

\item Harun, Z., Monty, J. P., Mathis, R. and Marusic, I. (2013), Pressure gradient effects on the large-scale structure of turbulent boundary layers. {\it J. Fluid Mech.}, Vol. 715, pp. 477--498.

\item Maciel, Y., Simens, M. P. and Gungor, A. G. (2017), Coherent structures in a nonequilibrium large-velocity-defect turbulent boundary layer. {\it Flow Turbul. Combust.}, Vol. 98, pp. 1--20.

\item Coleman, G. N., Rumsey, C. L. and Spalart, P. R. (2018), Numerical study of turbulent separation bubbles with varying pressure gradient and Reynolds number. {\it J. Fluid Mech.}, Vol. 847, pp. 28--70.

\end{References}
\end{document}